\documentclass{jkas}

\def\year{2015} % publication year
\def\month{October} % publication month
\def\volume{00} % journal volume
\def\issue{0} % journal issue
\def\beginpage{1} % first page of article
\def\endpage{99} % last page of article
\def\received{September xx, 2015} % date paper was received by JKAS
\def\accepted{October xx, 2015} % date of acceptance
\date{Received \received; accepted \accepted}

\usepackage{flushend} %% balance columns on last page
\title{
Monitoring of GAmma-ray Bright AGN: The Multi-frequency Polarization of the Flaring Blazar 3C~279\thanks{Part of a special issue on the Korean VLBI Network (KVN)}
}

\author[1,2]{Sincheol~Kang}
\author[1,2]{Sang-Sung~Lee}
\author[1]{Do-Young~Byun}
\affil[1]{Korea Astronomy and Space Science Institute, 776 Daedeok-daero, Yuseong, Daejeon 34055, Republic of Korea; \email{kang87@kasi.re.kr, sslee@kasi.re.kr, bdy@kasi.re.kr}}
\affil[2]{University of Science and Technology, 217 Gajeong-ro, Yuseong-gu, Daejeon 34113, Republic of Korea}
\def\corrauthor{
Sang-Sung Lee
}

\def\runningauthor{
Kang et al.
}

\def\runningtitle{
MOGABA: Multi-Frequency Polarization of 3C 279
}

\def\keywords{
galaxies: active --- galaxies: jets --- galaxies: magnetic fields --- polarization --- quasars: individual (3C~279)
}

\def\abstracttext{
We present results of long-term multi-wavelength polarization observations
of the powerful blazar 3C~279 after its $\gamma$-ray flare on 2013~December 20. 
We followed up this flare with single-dish polarization observations using two 21-m telescopes of the Korean VLBI Network. Observations carried out weekly from 2013~December~25 to 2015~January~11,
at 22~GHz, 43~GHz, 86~GHz simultaneously,
as part of the Monitoring Of GAmma-ray Bright AGN (MOGABA) program. We measured 3C~279 total flux densities of 22--34~Jy at 22~GHz, 15--28~Jy (43~GHz), and 10--21~Jy (86~GHz),
showing mild variability of $\leq 50\,\%$ over the period of our observations. The spectral index between 22~GHz and 86~GHz ranged
from $-0.13$ to $-0.36$. Linear polarization angles were
27$^{\circ}$--38$^{\circ}$, 30$^{\circ}$--42$^{\circ}$,
and 33$^{\circ}$--50$^{\circ}$ at 22~GHz, 43~GHz, and 86~GHz, respectively.
The degree of linear polarization was in the range of 6--12\,\%,
and slightly decreased with time at all frequencies.
We investigated Faraday rotation and depolarization of
the polarized emission at 22--86~GHz, and
found Faraday rotation measures (RM) of  $-300$ to $-1200$~rad~m$^{-2}$ between 22~GHz and 43~GHz, and $-800$ to $-5100$~rad~m$^{-2}$ between 43~GHz and 86~GHz.
The RM values follow a power law  with a mean power law index $a$ of $2.2$,
implying that the polarized emission at these frequencies
travels through a Faraday screen in or near the jet.
We conclude that
the regions emitting polarized radio emission may be different from the region
responsible for the 2013 December $\gamma$-ray flare and are maintained
by the dominant magnetic field perpendicular to the
direction of the radio jet at milliarcsecond scales.
}

\begin{document}
\jkashead %% set title, authors, abstract, etc.

%%%%%%%%%%%%%%%%%%%%%%%%%%%%%%%%%%%%%%%%%%%%%%%%%%%%%%%%%%%%%%%%%%%%%
%%% BEGIN MAIN TEXT HERE %%%%%%%%%%%%%%%%%%%%%%%%%%%%%%%%%%%%%%%%%%%%
%%%%%%%%%%%%%%%%%%%%%%%%%%%%%%%%%%%%%%%%%%%%%%%%%%%%%%%%%%%%%%%%%%%%%

\section{Introduction\label{sec:intro}}

Active Galactic Nuclei (AGN)
are among the most spectacular objects in the Universe. They
produce enormous luminosities ($\sim10^{42}$ to $\sim10^{48}$~erg~s$^{-1}$)
in compact regions (i.e., $\ll$1pc$^{3}$)
and radiate broadband continuum emission from
radio to $\gamma$-ray (up to TeV energies) with
often highly variable and polarized emission detected
in particular at radio, optical, and $\gamma$-ray~\citep[see, e.g.,][]{kro99}.
AGN are known to be powered by accretion of gas onto supermassive black holes
(SMBH, $M_{\rm BH}\approx10^{6}-10^{9}~M_{\odot}$),
launching relativistic outflows (jets)
perpendicular to the accretion disk plane into interstellar space \citep[see e.g.,][for a review]{boe+12}.
AGN with relativistic jets roughly aligned with our line of sight
(e.g., viewed within $\sim$20$^{\circ}$ from the jet axis)
are commonly called blazars.
Characteristic properties of blazars are strong continuum emission,
rapid variability, high polarization (at radio and optical),
and sometimes a lack of significant emission lines.
The high apparent Lorentz factors in the relativistic jets lead to amplification (boosting) of the continuum emission
by the fourth power of the relativistic Doppler factor~\citep{kra+12}. The narrow viewing angles and relativistic flows often lead to dramatic variations in total flux and polarization over the full spectrum.

Within last 7 years, $\gamma$-ray bursts (or flares)
have been reported from more than 100 AGN, including
3C~273\footnote{Atel \#2200: \url{www.astronomerstelegram.org/?read=2200}},
4C~+21.35\footnote{Atel \#2584: \url{www.astronomerstelegram.org/?read=2584}},
3C~454.3\footnote{Atel \#3041: \url{www.astronomerstelegram.org/?read=3041}},
and 1510-089\footnote{Atel \#3473: \url{www.astronomerstelegram.org/?read=3473}},
by the Large Area Telescope on the Fermi Gamma-ray Space
Telescope\footnote{\url{http://fermi.gsfc.nasa.gov}}.
Subsequent studies of the
correlation between
emission at $\gamma$-ray and other wavelengths have been aimed at
revealing, e.g., the origin and location of the $\gamma$-ray flares in AGN,
the dominating emission mechanisms of the high energy radiation
(including $\gamma$-rays),
and the role of magnetic fields in $\gamma$-ray flares.
For example, \citet{mar+10} compare the $\gamma$-ray light curve of PKS~1510-089 against other wavelengths using data
obtained while the source flared at $\gamma$-ray during 2009.0$\sim$2009.5.
They find that the $\gamma$-ray peak coincides with the optical peak.
They further find that the optical electric vector position angle (EVPA) rotated more than 700$^\circ$ during five days.
The authors conclude that the jets are threaded by helical magnetic fields and that the $\gamma$-ray flares and the optical variations are caused
by a single emission feature propagating in them.
Several $\gamma$-ray bright AGN are well-studied, including
the already mentioned PKS~1510-089 (cf. \citealt{mar+10,dam+11,ori+11}),
3C~454.3 (cf. \citealt{sas+10,weh+12}), and 3C~279 (cf. \citealt{lar+08,cha+08,abd+10,ale+14}).

The blazar
3C~279 is one of the brightest highly variable sources at radio to $\gamma$-ray wavelengths. Its multi-wavelength variability makes it one of
the most interesting objects for studying radio-loud AGN and correlations in variability between different wavelengths.
\cite{abd+10} investigated the  correlation between 3C~279 $\gamma$-ray flares and
the optical polarization angles using multi-wavelength data
obtained around MJD 54880
(2009 February 18)
%(18 February 2009)
when a $\gamma$-ray
flare was
detected by Fermi-LAT. They find that the optical EVPA rotates
by 208$^\circ$ after the $\gamma$-ray
flare,
while the flux density at radio
and millimeter wavelengths did not vary significantly, indicating that the region of
this event is optically thick at longer wavelengths (e.g., radio to millimeter).
On
2013 December 20,
another giant $\gamma$-ray
flare was
detected
in this source by Fermi-LAT\footnote{Atel \#5680: \url{www.astronomerstelegram.org/?read=5680}}.
\citet{hayashida+15}
investigated
a series of $\gamma$-ray flares
peaking on
2013 December 20 (MJD 56646),
2014 March 2 (MJD 56718),
and 2014 April 3 (MJD 56750).
Comparing the $\gamma$-ray, X-ray, optical, and radio light curves of 3C~279, they find that the X-ray flux increased by $\sim 50 \%$
and the soft X-ray flux and photon index followed a harder-when-brighter trend.
However, they were not able to find correlations between the $\gamma$-ray flares and other light curves at X-ray, optical (including polarization),
and radio (only total flux),
although the optical flux increased steadily after the $\gamma$-ray flare.
Moreover, the radio flux was stable compared to the $\gamma$-ray
and X-ray fluxes.

We conducted simultaneous multi-frequency (22~GHz, 43~GHz, and 86~GHz)
polarization observations using telescopes of the 
Korean Very Long Baseline Interferometry (VLBI) Network (KVN), 
in order to study the possible correlation between the $\gamma$-ray flare 
on 2013 December 20 and variation of chromatic radio polarization 
as well as radio flux density.
We describe the observations and the data reduction in Section~\ref{sec:obs and data}, and present the results, including the 3C~279 flux density,
the degree of linear polarization, and the linear polarization angle,
in Section~\ref{sec:results}.
Then, in Section~\ref{sec:discussion and conclusion}
we discuss the results focusing on investigating
the physical aspects of the chromatic radio polarization emission
from the $\gamma$-ray flaring AGN:
the spectral index,
the Faraday rotation measure,
and the factor of depolarization, and
provide our conclusions.
Our findings are summarized in Section~\ref{sec:summary}.

\section{Observations and Data Reduction\label{sec:obs and data}}

\subsection{Observations\label{subsec:obs}}

%%% TABLE %%%%%%%%%%%%%%%%%%%%%%%%%%%%%%%%%%%%%%%%%%%%%%%%%%%%%%%%%%%%%%%%%%%%%
\begin{table*}[t!]
\caption{Summary of our observations\label{tab1}}
\centering
\begin{tabular}{lcccc}
\toprule
~~~~~~~~~~~~~~~~Dates                                 & &               & $\nu$ (GHz)~:~HPBW (arcsec)~:~$\eta$  &           \\
\cmidrule(l{3em}r{3em}){3-5}
~~~~MJD~~~~~~~~~~~~~~~~~~~~~Calendar                  & &          TN         &            US        &            YS       \\
\midrule
56651-56991$^{\rm a}$~~~~~2013.12.25 -- 2014.11.30    & & 22.400~:~125~:~0.58 &      ---             & 43.122~:~63~:~0.64  \\
                                                      & & 43.100~:~61~:~0.60  &                      & 86.243~:~32~:~0.49  \\
56992-57022$^{\rm b}$~~~~~2014.12.01 -- 2014.12 31    & &         ---         & 22.400~:~124~:~0.61  & 43.122~:~63~:~0.64  \\
                                                      & &                     & 43.100~:~63~:~0.62   & 86.243~:~32~:~0.49  \\
57022-57033$^{\rm b}$~~~~~2015.01.01 -- 2015.01.11    & &         ---         & 22.400~:~124~:~0.63  & 43.122~:~63~:~0.63  \\
                                                      & &                     & 43.100~:~63~:~0.61   & 86.243~:~32~:~0.50  \\
\bottomrule
\end{tabular}
\tabnote{
a: epoch 1; b: epoch 2; in epoch 1, we did not conduct observations from 2013.6.14 to 2013.8.22 due to KVN summer maintenance.}
\end{table*}
%%%%%%%%%%%%%%%%%%%%%%%%%%%%%%%%%%%%%%%%%%%%%%%%%%%%%%%%%%%%%%%%%%%%%%%%%%%%%%%

Soon after the giant $\gamma$-ray
flare
of 3C~279
on 2013 December 20 (MJD 56646),
we started weekly single-dish
polarimetric observations of this radio-loud AGN between 2013 December 25 and 2015 March 10 using KVN radio telescopes. Details of the observations are described in Table~\ref{tab1}.

The observations were conducted as part of a KVN monitoring program called
MOnitoring Of GAmma-ray Bright-AGN (MOGABA; cf. \citealt{lee+13}).
The KVN is a VLBI network dedicated to mm-wavelengths and it consists of three 21-m
radio telescopes in Korea~\citep{lee+14}.
Each telescope can observe at four frequency bands (22~GHz, 43~GHz, 86~GHz and 129~GHz), simultaneously, in single polarization, or at any two out of these four bands in dual circular polarization
(Left-handed Circular Polarization, LCP and Right-handed
Circular Polarization, RCP).
For details on the KVN single-dish systems, see \citet{lee+11}.
For our single-dish
polarimetric observations of 3C~279
at 22~GHz, 43~GHz, and 86~GHz, we used two KVN telescopes,
since  KVN telescopes have dual-polarization capability and support single-dish (non-VLBI) observing.

We observed (quasi-)simultaneously at three frequencies (22.400~GHz, 43.122~GHz, and 86.243~GHz) and used the KVN digital spectrometer with 4096 spectral channels across a bandwidth of 512 MHz, corresponding to a channel spacing of 125~kHz. For accurate polarimetry we also measured cross-polarization spectra. We aimed to study both intra-day and long-term variations of 3C~279, thus observations were conducted once a week (interrupted by KVN maintenance), with each observing session containing of 6 or 7 scans on 3C~279, and one scan on each calibrator, such as the Crab nebula, planets, and 3C~286.

Each observation session included multiple polarization observations of 3C~279 and calibrators. Each polarization observation of 3C~279 was composed of 8 sets of position-switching cycles, each with 16 on-off switching measurements towards the source and a reference sky patch, with a typical on-source integration time of 348~s (=$8\times16\times3$~s, where the on-source integration time for an on-off switching is 3~s).
The duration of each polarization observation was about 40 minutes, including the antenna slewing time.
We minimized telescope movement and calibrated the polynomial drifts of receiver output power, by conducting on-off switching measurements
following a switching cycle
OFF--ON--ON--OFF--ON--OFF--OFF--ON--ON--OFF--OFF--ON--OFF--ON--ON--OFF--ON--OFF--OFF--ON--OFF--ON--ON--OFF--OFF--ON--ON--OFF--ON--OFF--OFF--ON,
as shown in Figure E.1 of \cite{man00}\footnote{\url{http://aro.as.arizona.edu/12m_docs/12m_userman.pdf}}.
The polynomial drifts of receiver output power are mainly due to the receiver gain variations
and the atmospheric fluctuations on timescales from a few seconds to a few hundreds seconds

Before each observing session, we corrected antenna pointing offsets
and measured the total flux density (i.e., Stokes I) using cross-scan observations of the source.
All observations followed
antenna gain calibration measurements.
To track changes in atmospheric opacity, once between every two polarization observations we measured the system temperature at eight elevations,
18.21$^{\circ}$, 20.17$^{\circ}$, 22.62$^{\circ}$, 25.77$^{\circ}$,
30.00$^{\circ}$, 36.03$^{\circ}$, 45.58$^{\circ}$, and 65.38$^{\circ}$
(so-called sky tipping curve measurements).
The typical optical depths during our observations were 0.06--0.08, 0.16--0.18,
and 0.12--0.15 in the 22~GHz, 43~GHz, and 86~GHz bands, respectively.

To calibrate the polarization angle, the Crab nebula was observed in
each session (i.e., about once per seven days).
The polarization angle of the Crab nebula at the KVN operating frequencies is
well known and stable in time~\citep{fh79,aum+10}.
Planets, such as Venus, Jupiter, or Mars were observed in each session for calibrating the instrumental polarization, based on the assumption that their emission is unpolarized.
To ensure the reliability of the polarization calibration,
the standard polarization calibrator 3C~286 was observed once per session
as well.
We calibrated the absolute flux density scale with antenna gain
measurements using planets and the antenna gain--elevation curve~\citep{lee+11}. All observations were made at elevations between
30$^{\circ}$ and 70$^{\circ}$, since according to the KVN status
report of 2014, the antenna gains are stable
in this elevation range\footnote{\url{http://kvn.kasi.re.kr/status_report_2014}}.

\subsection{Data Reduction\label{subsec:data}}

%%% FIGURE %%%%%%%%%%%%%%%%%%%%%%%%%%%%%%%%%%%%%%%%%%%%%%%%%%%%%%%%%%%%%%%%%%%%
\begin{figure*}[t!]
\centering
\includegraphics[trim=5mm 5mm 20mm 10mm, clip, width=155mm]{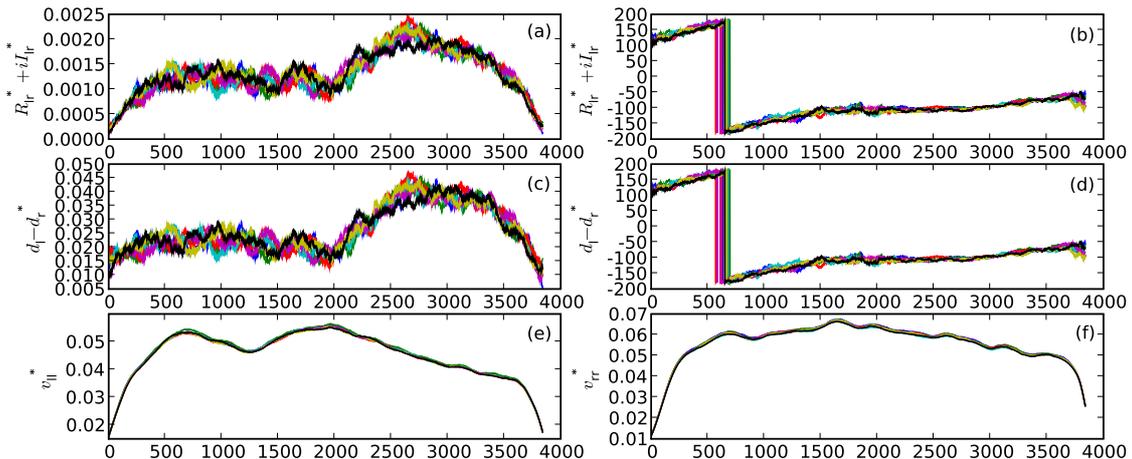}
\caption{
Spectrometer outputs for Jupiter.
(a) and (b) are the amplitude and phase of $R_{\rm rl^*}+iI_{\rm rl^*}$, respectively,
(c) and (d) are the amplitude and phase of $d_{\rm l}-d_{\rm r}^*$, respectively.
(e) is $v_{\rm ll^*}$, and (f) is $v_{\rm rr^*}$.
Colors indicate the 8 sets of measurements.
\label{fig:unpol}}
\end{figure*}
%%%%%%%%%%%%%%%%%%%%%%%%%%%%%%%%%%%%%%%%%%%%%%%%%%%%%%%%%%%%%%%%%%%%%%%%%%%%%%%

%%% FIGURE %%%%%%%%%%%%%%%%%%%%%%%%%%%%%%%%%%%%%%%%%%%%%%%%%%%%%%%%%%%%%%%%%%%%
\begin{figure*}[t!]
\centering
\includegraphics[trim=5mm 5mm 20mm 10mm, clip, width=155mm]{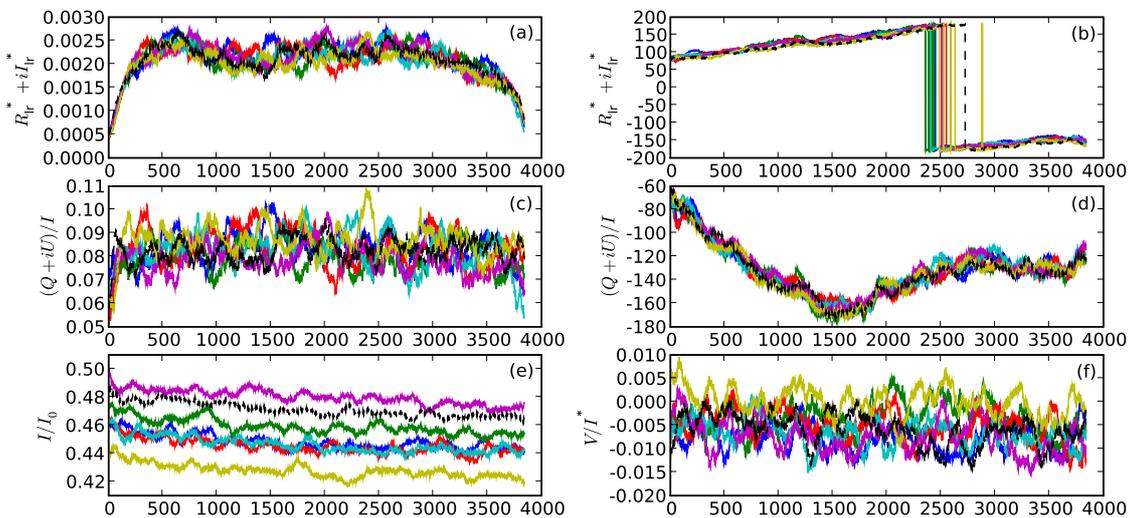}
\caption{Spectrometer outputs for 3C~279.
(a) and (b) are the amplitude and phase of $R_{\rm rl^*}+iI_{\rm rl^*}$,
(c) and (d) are the amplitude and phase of $\frac{Q+iU}{I}$, respectively.
(e) is $\frac{I}{I_{\rm o}}$, and (f) is $\frac{V}{I}$.
Colors indicate the 8 sets of measurements.
\label{fig:source}}
\end{figure*}
%%%%%%%%%%%%%%%%%%%%%%%%%%%%%%%%%%%%%%%%%%%%%%%%%%%%%%%%%%%%%%%%%%%%%%%%%%%%%%%

We processed the polarimetric observations
using the data processing software developed by Byun et al. (in prep.),
which we briefly describe in this section.
The KVN digital spectrometer backend used for single-dish polarimetric observations outputs two single-polarization spectra
($v_{\rm ll^*}$ and $v_{\rm rr^*}$),
and one complex cross-polarization spectrum
($v_{\rm lr^*}$ or $v_{\rm rl^*}$, where $v_{\rm rl^*}$
is the complex conjugate of $v_{\rm lr^*}$)
separated into real and imaginary parts
($R_{\rm rl^*}$ and $I_{\rm rl^*}$).
All four Stokes parameters,
$I$, $Q$, $U$, and $V$,
can be estimated from these four polarization outputs
using the following equations~\citep{Sault et al. (1996)}:
%%%%%%%%%%%%%%%%%%%%%%%%%%%%%%%%%%%%%%%%%%%%%%%%%%%%%%%%%%%%%%%%%%%%%%%%%%%%%%
\begin{eqnarray}
\label{eq:stokes parameters}
{v_{\rm ll^*}} &=& \frac{1}{2}g_{\rm l}g_{\rm l}^*\left(I+V\right)        \\
{v_{\rm lr^*}} &=& \frac{1}{2}g_{\rm l}g_{\rm r}^*\left[\left(d_{\rm l}-d_{\rm r}^*\right)I+e^{-2i\chi_{\rm PA}}\left(Q+iU\right)\right]        \\
{v_{\rm rl^*}} &=& \frac{1}{2}g_{\rm r}g_{\rm l}^*\left[\left(d_{\rm l}^*-d_{\rm r}\right)I+e^{2i\chi_{\rm PA}}\left(Q-iU\right)\right]        \\
{v_{\rm rr^*}} &=& \frac{1}{2}g_{\rm r}g_{\rm r}^*\left(I-V\right)
\end{eqnarray}
%%%%%%%%%%%%%%%%%%%%%%%%%%%%%%%%%%%%%%%%%%%%%%%%%%%%%%%%%%%%%%%%%%%%%%%%%%%%%%
where $v_{\rm ll^*}$, $v_{\rm lr^*}$, $v_{\rm rl^*}$ and $v_{\rm rr^*}$
are spectrometer
outputs of circular feed horns, the subscripts $l$, $r$, $l^*$, and $r^*$ indicate
left-handed and right-handed circular polarizations and
their complex conjugates, respectively,
$g_{\rm l}$ and $g_{\rm r}$ are complex gain factors,
$d_{\rm l}$ and $d_{\rm r}$ are complex terms describing cross-polarization leakage ($d_{\rm l}$: LCP to RCP, $d_{\rm r}$: RCP to LCP), i.e., the D-term,
and $\chi_{\rm PA}$ is the parallactic angle.
With the polarization outputs from the KVN spectrometer,
we were able to determine the complex spectra of
so-called combined-D-term $d_{\rm l}-d_{\rm r}^*$ (or $d_{\rm r}-d_{\rm l}^*$)
by observing unpolarized sources such as Jupiter, Venus, or Mars,
as shown in Figure~\ref{fig:unpol}a for the amplitude and
in Figure~\ref{fig:unpol}b for the phase,
assuming the Stokes parameters $V=Q=U=0$ for the unpolarized sources.
Then, we estimated the four Stokes parameters,
i.e., {$I$, $Q$, $U$}, and {$V$}
for the target sources.

\begin{figure*}[!t]
\centering
\includegraphics[trim=30mm 7mm 40mm 20mm, clip, width=174mm]{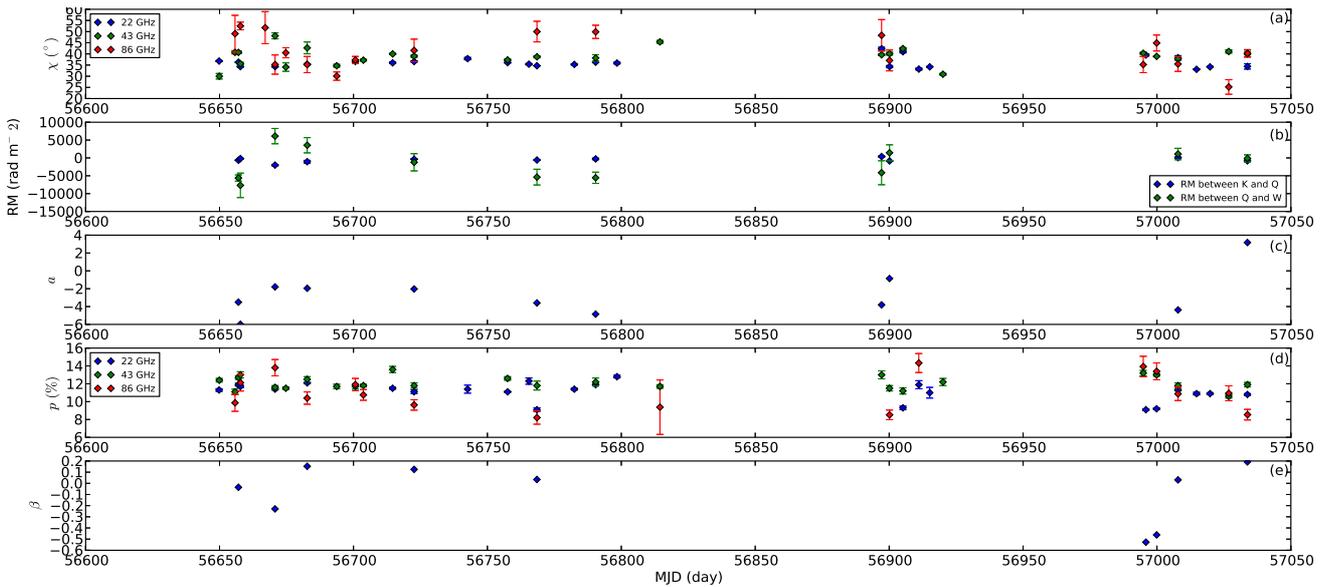}
\caption{Light curves of a comparison source 3C~286
for (a) the linear polarization angle ($^{\circ}$),
(b) the Faraday rotation measure RM (rad~m$^{-2}$),
(c) the $a$ index,
(d) the degree of linear polarization (\%),
and (e) the depolarization index $\beta$,
at 22~GHz (red), 43~GHz (green),
and 86~GHz (blue) from 2013 December to 2015 March.
3C~286 is much fainter at 86~GHz than at 22 and 43~GHz,
so the uncertainty of the polarization observations
due to thermal noise is larger.
The $a$ index and the depolarization index $\beta$
are derived when the polarization measurements at three frequency bands
are available.
\label{fig:3c286}}
\end{figure*}

From the estimated Stokes parameters, we obtained
the total flux density, $I$,
the degree of linear polarization, $p=\sqrt{Q^{2}+U^{2}} / I$,
and the polarization angle
$\chi = (1/2)\times\arctan(U/Q)$,
as shown in Figure~\ref{fig:source}.
The estimated polarization angle is converted to the intrinsic
polarization angle by comparing
the estimated polarization angle against the known angle of a primary angle calibrator, like the Crab nebula.
We used $\chi=154^{\circ}\pm2^{\circ}$ for the Crab nebula at 22--86~GHz
bands~\citep{fh79,aum+10}.
We evaluated the polarization observations
by calibrating the linear polarization angles of 3C~286 when available.
This target has stable polarization angles of
$35\pm0.2^{\circ}$ at 23~GHz, and $35.8\pm0.1^{\circ}$ at 45~GHz~\citep{pb13}.
The polarization angles of 3C~286 obtained
from our KVN observations at 22~GHz and 43~GHz are
shown in Figure~\ref{fig:3c286}.
They are largely consistent with those of \citeauthor{pb13}.

We flagged 3C~279 polarization measurements in a session
when the polarization angle of 3C~286 at 22~GHz or 43~GHz deviated from
its intrinsic value or when the polarization angle uncertainty of 3C~279
was larger than the expected value.
The rms uncertainties of the linear polarization observations
due to the thermal noise were
about 10~mJy and 15~mJy at 22~GHz and 43~GHz, respectively.
The systematic error of the polarization angle measurements
was  $\sim2^\circ$ at both frequencies due to the uncertainty of
the polarization angle of the Crab nebula.
The typical instrumental polarization leakage of the KVN system
is $<5\%$ at both frequencies, e.g., as shown in Figure~\ref{fig:unpol}c.

We also obtained the total flux density of 3C~279 from the cross scan
observations taking into account the antenna pointing corrections
as
$${T_{\rm A,az}^{*}}={T_{\rm A,az,o}^{*}}\times\exp\left(4\ln2\frac{EL_{\rm off}}{HPBW}\right)$$
and
$${T_{\rm A,el}^{*}}={T_{\rm A,el,o}}\times\exp\left(4\ln2\frac{AZ_{\rm off}}{HPBW}\right) , $$
where $T_{\rm A,az,o}$ and $T_{\rm A,el,o}$ are the
peak values of the observed antenna temperatures for azimuth and elevation scans,
respectively, $HPBW$ is half power beam width of a KVN single-dish antenna,
and $EL_{\rm off}$ and $EL_{\rm off}$ are the pointing offsets in
the azimuth and elevation directions, respectively.
We obtained the averaged peak value $T_{\rm A}$ and converted it to
the source flux density as
$S_{\nu} = 2kT / (\eta A_{\rm g})$,
where $k$ is the Boltzmann constant, and $\eta$ and $A_{\rm g}$ are the aperture efficiency and area of a KVN single-dish antenna.
For each observing epoch we adopted appropriate values for $HPBW$ and $A_{\rm g}$ (see Table~\ref{tab1}) from the annual KVN status reports.

\section{Results\label{sec:results}}

\begin{table*}[t!]
\caption{Results of multi-frequency polarization observations for 3C~279. Uncertainties are $1\sigma$. \label{tab2}}
\centering
\resizebox{175mm}{!}{%% not presented correctly in some DVI viewers -- look at PDF!
\begin{tabular}{lccccccccc}
\toprule
         &       & 22~GHz&       &       & 43~GHz &        &       & 86~GHz &        \\
\cmidrule(l{1em}r{1em}){2-4}\cmidrule(l{1em}r{1em}){5-7}\cmidrule(l{1em}r{1em}){8-10}
MJD      &$S_{\nu}$~(Jy) &$\chi~(^\circ)$& $p$~(\%)  &$S_{\nu}$~(Jy)&$\chi~(^\circ)$& $p$~(\%)  &$S_{\nu}$~(Jy)&$\chi~(^\circ)$& $p$~(\%)      \\
\midrule
56649.86 & 29.4$\pm$0.01 & 34.8$\pm$0.07 & 9.2$\pm$0.02 & 25.7$\pm$0.01 & 31.6$\pm$0.11 & 10.0$\pm$0.01 & 18.6$\pm$0.02 & 36.4$\pm$0.21 & 9.6$\pm$0.05 \\
56654.82 &               &               &              & 26.1$\pm$0.01 & 36.3$\pm$0.14 & 9.5$\pm$0.01 & 20.4$\pm$0.02 & 42.6$\pm$0.17 & 9.4$\pm$0.06 \\
56655.84 &               &               &              & 24.6$\pm$0.01 & 36.8$\pm$0.06 & 9.8$\pm$0.01 & 16.4$\pm$0.02 & 41.6$\pm$0.64 & 10.4$\pm$0.05 \\
56656.81 & 29.5$\pm$0.01 & 32.4$\pm$0.05 & 9.6$\pm$0.02 & 26.2$\pm$0.01 & 37.9$\pm$0.35 & 9.7$\pm$0.02 & 20.5$\pm$0.02 & 41.1$\pm$0.21 & 10.2$\pm$0.03 \\
56657.83 & 30.3$\pm$0.01 & 33.7$\pm$0.04 & 9.5$\pm$0.02 & 25.7$\pm$0.01 & 36.2$\pm$0.08 & 9.7$\pm$0.02 & 14.8$\pm$0.02 & 41.8$\pm$0.29 & 10.0$\pm$0.05 \\
56670.77 & 30.9$\pm$0.01 & 30.3$\pm$0.01 & 9.8$\pm$0.01 & 26.3$\pm$0.01 & 35.6$\pm$0.17 & 10.0$\pm$0.02 & 17.3$\pm$0.02 & 39.4$\pm$0.18 & 10.7$\pm$0.05 \\
56674.77 &               &               &              & 26.2$\pm$0.01 & 32.3$\pm$0.17 & 10.1$\pm$0.01 & 19.3$\pm$0.02 & 34.9$\pm$0.15 & 10.6$\pm$0.04 \\
56682.78 & 30.9$\pm$0.01 & 29.5$\pm$0.24 & 9.6$\pm$0.02 & 26.5$\pm$0.01 & 31.7$\pm$0.07 & 9.5$\pm$0.01 & 18.6$\pm$0.01 & 36.4$\pm$0.15 & 9.7$\pm$0.04 \\
56693.76 &               &               &              & 27.8$\pm$0.01 & 30.9$\pm$0.06 & 9.6$\pm$0.01 & 20.5$\pm$0.02 & 33.8$\pm$0.15 & 10.1$\pm$0.03 \\
56700.77 &               &               &              & 27.8$\pm$0.01 & 31.8$\pm$0.08 & 9.8$\pm$0.02 & 20.1$\pm$0.02 & 35.1$\pm$0.11 & 10.5$\pm$0.03 \\
56703.72 &               &               &              & 27.9$\pm$0.01 & 30.5$\pm$0.10 & 9.8$\pm$0.01 & 17.7$\pm$0.02 & 35.8$\pm$0.34 & 10.4$\pm$0.06 \\
56714.65 & 33.0$\pm$0.01 & 29.9$\pm$0.04 & 9.4$\pm$0.02 & 28.0$\pm$0.01 & 34.4$\pm$0.15 & 10.0$\pm$0.01 & 19.9$\pm$0.02 & 39.5$\pm$0.22 & 11.0$\pm$0.03 \\
56722.66 & 31.7$\pm$0.01 & 32.1$\pm$0.02 & 9.9$\pm$0.01 & 27.0$\pm$0.01 & 34.7$\pm$0.05 & 10.5$\pm$0.01 & 19.4$\pm$0.02 & 40.0$\pm$0.07 & 11.8$\pm$0.03 \\
56742.63 & 32.2$\pm$0.01 & 28.2$\pm$0.01 & 9.9$\pm$0.02 &      &      &      &         &           &     \\
56757.55 & 29.2$\pm$0.01 & 34.4$\pm$0.06 & 9.3$\pm$0.02 & 24.9$\pm$0.01 & 36.8$\pm$0.06 & 9.9$\pm$0.01 & 16.2$\pm$0.02 & 43.9$\pm$0.19 & 10.5$\pm$0.08 \\
56765.51 & 29.3$\pm$0.02 & 33.6$\pm$0.10 & 9.0$\pm$0.02 &      &      &      &         &           &     \\
56768.52 & 29.2$\pm$0.02 & 31.9$\pm$0.08 & 8.9$\pm$0.02 & 24.2$\pm$0.01 & 37.4$\pm$0.09 & 9.8$\pm$0.01 & 18.2$\pm$0.02 & 44.4$\pm$0.20 & 11.0$\pm$0.06 \\
56782.47 & 28.6$\pm$0.01 & 32.4$\pm$0.54 & 8.9$\pm$0.04 &      &      &      &         &           &     \\
56790.44 & 28.4$\pm$0.01 & 35.1$\pm$0.08 & 8.6$\pm$0.01 & 23.5$\pm$0.01 & 37.7$\pm$0.11 & 9.3$\pm$0.03 & 15.1$\pm$0.02 & 44.3$\pm$0.27 & 10.3$\pm$0.13 \\
56798.43 & 28.9$\pm$0.01 & 37.6$\pm$0.57 & 8.5$\pm$0.04 &      &      &      &         &           &     \\
56814.45 &               &               &              & 22.6$\pm$0.01 & 41.1$\pm$0.34 & 9.0$\pm$0.02 & 16.1$\pm$0.03 & 41.1$\pm$0.59 & 9.4$\pm$0.09  \\
56897.20 & 33.1$\pm$0.09 & 31.8$\pm$0.11 & 6.8$\pm$0.20 & 19.2$\pm$0.01 & 38.6$\pm$0.22 & 7.6$\pm$0.05 & 12.0$\pm$0.03 & 45.0$\pm$0.56 & 7.7$\pm$0.35 \\
56900.17 & 29.6$\pm$0.08 & 30.5$\pm$0.53 & 6.4$\pm$0.08 & 20.3$\pm$0.01 & 36.8$\pm$0.19 & 7.5$\pm$0.05 & 15.2$\pm$0.03 & 43.1$\pm$1.10 & 8.6$\pm$0.33 \\
56905.19 & 27.4$\pm$0.05 & 34.2$\pm$0.17 & 6.3$\pm$0.10 & 17.2$\pm$0.05 & 42.0$\pm$0.16 & 8.3$\pm$0.16 &  &  &  \\
56911.13 & 29.7$\pm$0.05 & 27.9$\pm$0.20 & 6.5$\pm$0.05 &      &      &      &      &    &          \\
56915.17 & 27.3$\pm$0.08 & 28.2$\pm$0.18 & 6.0$\pm$0.06 &      &      &      &      &    &          \\
56920.13 &               &               &              & 20.9$\pm$0.01 & 33.5$\pm$0.36 & 6.4$\pm$0.05 & 16.6$\pm$0.02 & 43.8$\pm$1.31 & 8.1$\pm$0.16 \\
56994.90 & 24.6$\pm$0.01 & 32.0$\pm$0.16 & 7.7$\pm$0.07 & 15.4$\pm$0.01 & 38.5$\pm$0.06 & 8.4$\pm$0.03 & 10.2$\pm$0.01 & 49.2$\pm$0.49 & 9.2$\pm$0.09 \\
56999.90 &               &               &              & 18.6$\pm$0.01 & 38.8$\pm$0.12 & 7.9$\pm$0.02 & 11.8$\pm$0.01 & 48.0$\pm$0.24 & 8.7$\pm$0.05 \\
57007.88 & 22.8$\pm$0.01 & 35.2$\pm$0.57 & 7.2$\pm$0.05 & 18.9$\pm$0.01 & 40.0$\pm$0.06 & 7.9$\pm$0.01 & 12.5$\pm$0.01 & 46.6$\pm$0.09 & 9.1$\pm$0.07 \\
57014.80 & 23.8$\pm$0.01 & 30.8$\pm$0.09 & 6.7$\pm$0.02 &      &      &      &         &           &     \\
57019.83 & 23.6$\pm$0.01 & 32.8$\pm$0.14 & 6.6$\pm$0.01 &  &      &      &         &           &     \\
57026.82 &               &               &              & 18.9$\pm$0.01 & 42.1$\pm$0.23 & 7.1$\pm$0.02 & 14.2$\pm$0.02 & 43.8$\pm$0.21 & 9.2$\pm$0.05 \\
57033.78 & 23.0$\pm$0.01 & 32.3$\pm$0.28 & 7.2$\pm$0.02 & 18.7$\pm$0.01 & 40.4$\pm$0.21 & 7.5$\pm$0.02 & 13.3$\pm$0.02 & 46.8$\pm$0.18 & 8.8$\pm$0.08 \\
57040.78 & 24.0$\pm$0.01 & 31.0$\pm$0.45 & 8.4$\pm$0.09 &      &      &      &      &      &      \\
\bottomrule
\end{tabular}
}
\vskip1em
\end{table*}

%%% FIGURE %%%%%%%%%%%%%%%%%%%%%%%%%%%%%%%%%%%%%%%%%%%%%%%%%%%%%%%%%%%%%%%%%%%%
%%% FIGURE %%%%%%%%%%%%%%%%%%%%%%%%%%%%%%%%%%%%%%%%%%%%%%%%%%%%%%%%%%%%%%%%%%%%
\begin{figure*}[!t]
\centering
\includegraphics[trim=10mm 5mm 20mm 15mm, clip, width=174mm]{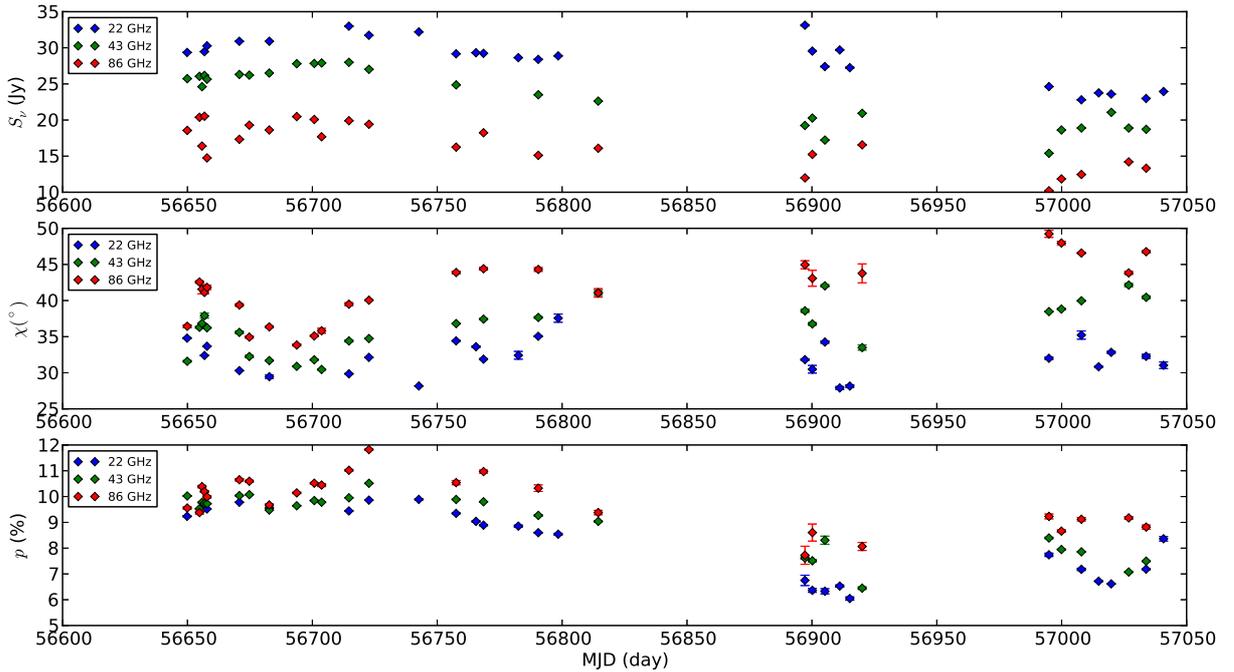}
\caption{Light curves of 3C~279
for the total flux density (top) measured with the cross-scan observations,
the linear polarization angle ($^{\circ}$) (middle),
and the degree of linear polarization (\%) (bottom),
at 22~GHz (red), 43~GHz (green),
and 86~GHz (blue) from 2013 December to 2015 March.
\label{fig:light curve}}
\end{figure*}
%%%%%%%%%%%%%%%%%%%%%%%%%%%%%%%%%%%%%%%%%%%%%%%%%%%%%%%%%%%%%%%%%%%%%%%%%%%%%%%

%%% FIGURE %%%%%%%%%%%%%%%%%%%%%%%%%%%%%%%%%%%%%%%%%%%%%%%%%%%%%%%%%%%%%%%%%%%%
\begin{figure*}[!t]
\centering
\includegraphics[trim=10mm 5mm 20mm 15mm, clip, width=174mm]{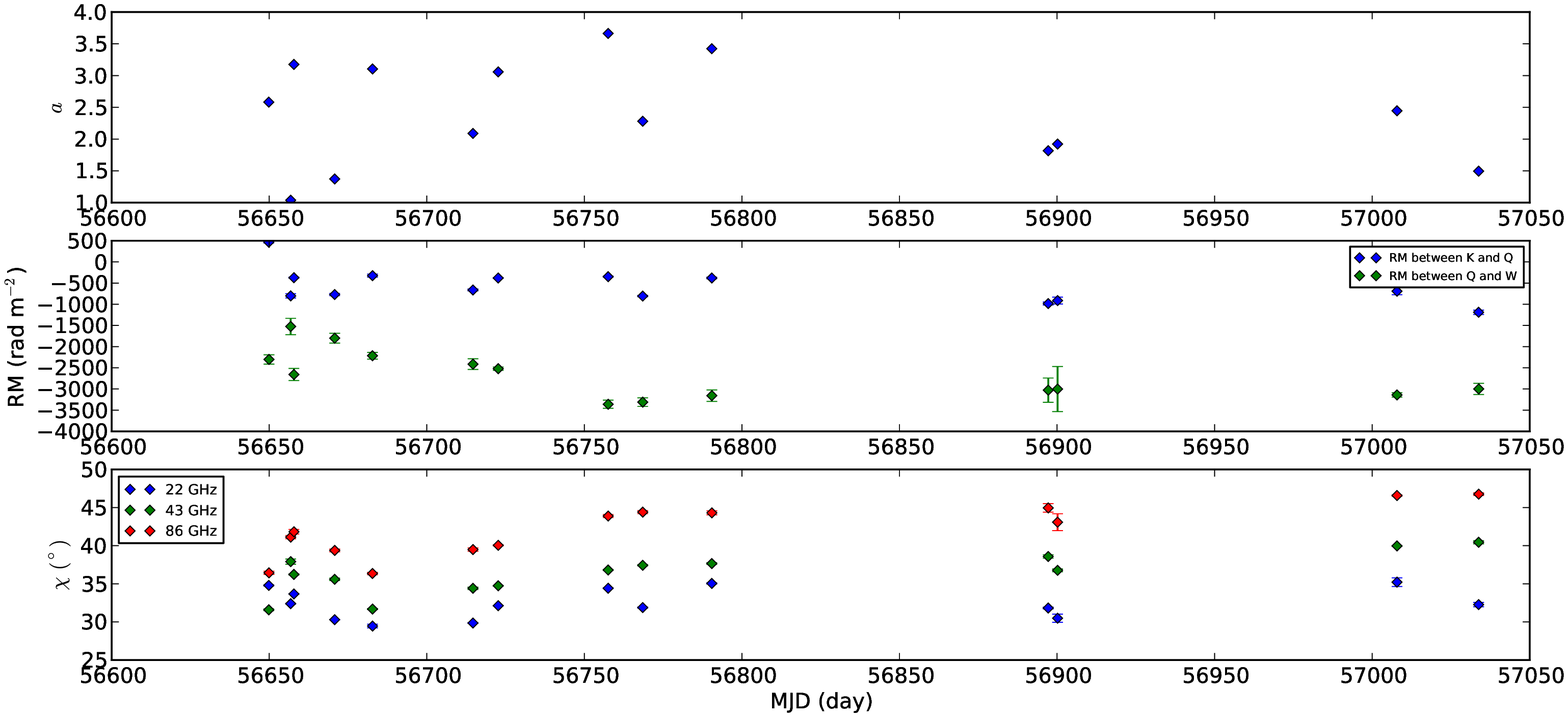}
\caption{
Plots of index
$a$ (top),
the Faraday rotation measure RM (rad~m$^{-2}$) (middle),
and the linear polarization angle ($^{\circ}$) (bottom) of 3C~279
at 22~GHz (red), 43~GHz (green), and 86~GHz (blue).
Measurements are shown only when the measurements at
three frequency bands are available for estimating the index $a$ in the relation $|{\rm RM}(\nu)|\propto\nu^{a}$.
\label{fig:rotation_measure}}
\end{figure*}
%%%%%%%%%%%%%%%%%%%%%%%%%%%%%%%%%%%%%%%%%%%%%%%%%%%%%%%%%%%%%%%%%%%%%%%%%%%%%%%

%%% FIGURE %%%%%%%%%%%%%%%%%%%%%%%%%%%%%%%%%%%%%%%%%%%%%%%%%%%%%%%%%%%%%%%%%%%%
\begin{figure*}[!t]
\centering
\includegraphics[trim=10mm 1mm 20mm 10mm, clip, width=174mm]{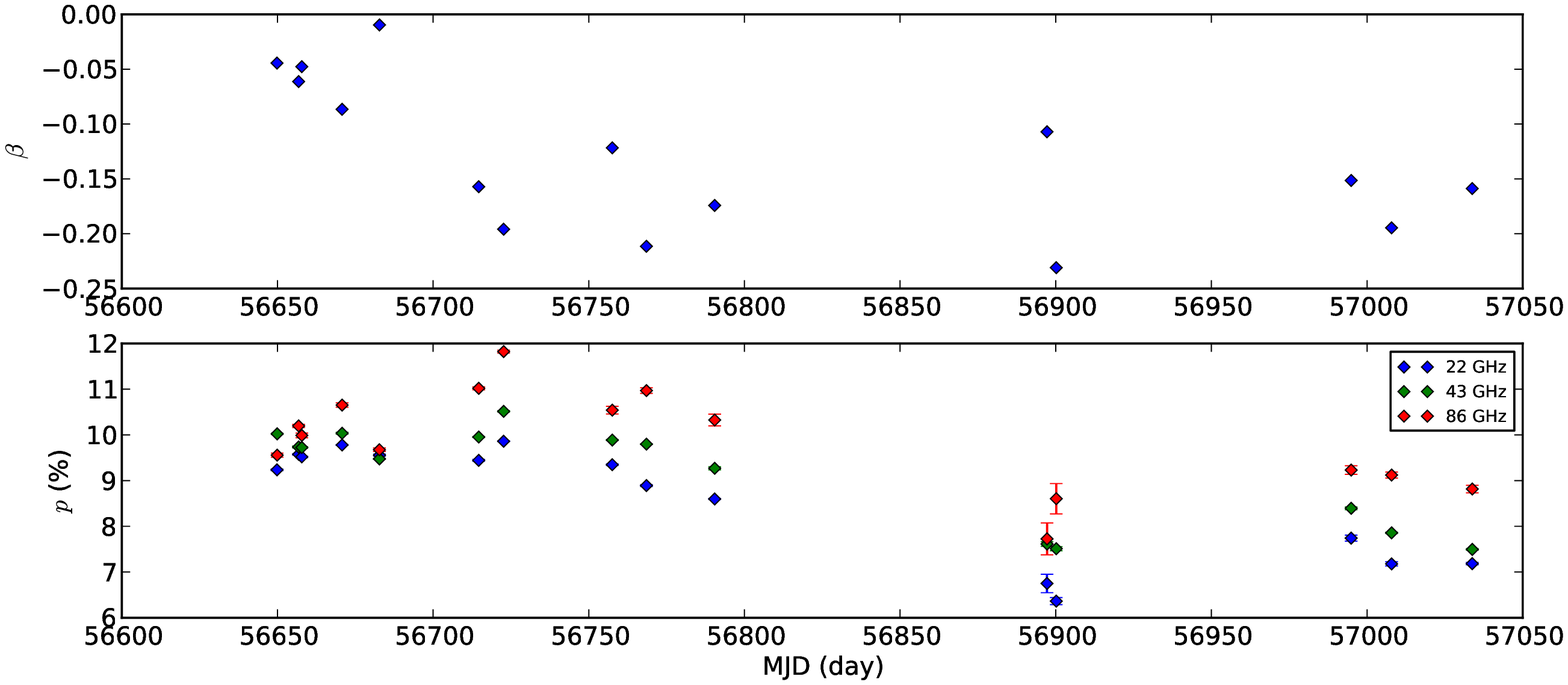}
%\caption{An example figure, from \citet{park2012}.\label{fig:jkasfig1}}
\caption{
Plots of depolarization index $\beta$ (top)
and the degree of linear polarization (bottom) of 3C~279
at 22~GHz (red), 43~GHz (green), and 86~GHz (blue).
Measurements are shown only when the measurements at
three frequency bands are available for estimating $\beta$ index.
The green solid line is the best fit result of the linear model to the data.
\label{fig:depolarization}}
\end{figure*}
%%%%%%%%%%%%%%%%%%%%%%%%%%%%%%%%%%%%%%%%%%%%%%%%%%%%%%%%%%%%%%%%%%%%%%%%%%%%%%%

Our multi-frequency simultaneous polarization
monitoring observations between 2013 December 25 and 2015 January 11
obtained 3C~279 total flux densities $S_{\nu}$ of 22--34~Jy, 15--28~Jy,
and 10--21~Jy at 22~GHz, 43~GHz, and 86~GHz, respectively.
The respective light curves are shown in Figure~\ref{fig:light curve} (top)
and are summarized in Table~\ref{tab2}.

The flux densities exhibit a mild global decrease with time of $<$50\%, with no prominent flares
over a period of about 390 days (MJD 56650--57040),
and were almost constant or slightly increasing over a period of about 80 days
(MJD 56650--57040).

The spectral index, $\alpha$ ($S_{\nu}\propto\nu^{\alpha}$),
between the 22~GHz and 86~GHz bands ranges from  $-0.13$ to $-0.36$,
implying that 3C~279 is a flat spectrum source. Its spectral index showed mild variability for about a year
after the 2013 December giant $\gamma$-ray
flare.

The linear polarization angle $\chi$ of 3C~279 is shown in Figure~\ref{fig:light curve} (middle), and is in the range of
27$^{\circ}$--38$^{\circ}$,
30$^{\circ}$--42$^{\circ}$, and
33$^{\circ}$--50$^{\circ}$
at 22~GHz, 43~GHz, and 86~GHz, respectively.
After the giant $\gamma$-ray flare, the polarization angle rotated
by $\sim$10$^{\circ}$ on a timescale of $\sim$50 days.

Faraday rotation scales
with the square of the observing wavelength, $\lambda$, like
$\chi_{\rm obs} = \chi_{\rm int} + \textup{RM} \lambda^2$. Using the polarization angles observed at 22--86~GHz,
we estimate the rotation measure, $\textup{RM}$, of the Faraday rotation of the polarized radio emission from 3C~279. The resulting $\textup{RMs}$ are shown in Figure~\ref{fig:rotation_measure} (middle). We find rotation measures
in the range from $-300$ to $-1200$~rad~m$^{-2}$ between the 22~GHz and 43~GHz bands,
and from $-800$ to $-5100$~rad~m$^{-2}$ between the 43~GHz and 86~GHz bands, respectively.
An exception was the first measurement on MJD 56650
with a positive RM of about $+500$~rad~m$^{-2}$ between the 22~GHz and 43~GHz bands,
leading to a RM change of about $-1300$~rad~m$^{-2}$ until the next
measurement on MJD 56657.

The degree of linear polarization, $p$, of 3C~279 measured in the 22--86~GHz bands
is in the range of 6\,\%--12\,\%.
It exhibits a slightly decreasing trend with time over the whole period of observations.
About 70 days after the giant $\gamma$-ray flare,
the degree of linear polarization at 86~GHz showed a short-term increase
from $\sim 10\%$ to $\sim 12\%$, whereas the polarization fraction at other frequencies
remained almost constant.
After MJD 56750 the degree of linear polarization at all frequency bands
started to decrease with time, and remained at low levels of around 6\,\%--9\,\%, although there are observation gaps due to KVN maintenance.

\section{Discussion and Conclusions\label{sec:discussion and conclusion}}

The multi-frequency light curves in total flux density
of 3C~279 at the 22~GHz, 43~GHz, and 86~GHz bands show
a mild global decrease with time of $<$50\%
with no prominent flares
over a period of $\sim$390 days (MJD 56650--57040),
and are almost constant or slightly increasing
over a period of $\sim$80 days (MJD 56650--56730),
maintaining the characteristics of a flat-spectrum radio quasar
with a spectral index of
$\alpha=-0.13\sim-0.36$.
We do not find a flaring
brightening by a factor of $>5$, which would be expected
for a radio counterpart to the strong $\gamma$-ray flare
detected in 2013 December (MJD 56650)~\citep[see Figure 1 in][]{hayashida+15}.
The lack of such a radio flare
may imply that
the radio and $\gamma$-ray emission are generated by different populations of electrons,
and at different locations, as indicated by \cite{hayashida+15} for the X-ray, optical,
and millimeter (230~GHz) bands.
A similar report on the poor correlation of the light curves
at radio (5, 15, 37, and 230~GHz)
and $\gamma$-ray
after the giant $\gamma$-ray flares in 2009 was given by \cite{abd+10},
indicating that the emission region responsible for
the $\gamma$-ray flares is optically thick at radio wavelengths.
It is possible that we have missed radio counterparts to the $\gamma$-ray flare,
if these occurred right after the $\gamma$-ray peak on 2013 December 20,
as there was a 4-day delay between the $\gamma$-ray flare
and the start of our observations.
%If, however, there were radio counter parts within 4 days
%after the gamma-ray flare,
%noting that our observations started 4 days
%after the peak of the gamma-ray flare on 2013 December 20,
%we could have missed them.
However, this possibility can be excluded by 
the 230~GHz light curve obtained before and after the gamma-ray peak
as reported by \cite{hayashida+15}, where
no prominent flare is detected,
assuming that any gamma-ray flare optically thin at 22-86~GHz
should be most likely optically thin at 230~GHz as well.
Therefore, we suggest that the 2013 December $\gamma$-ray
flare
of 3C~279 originated in an emission region
that is optically thick at radio frequencies of 22-86~GHz,
or may be generated by different populations of electrons
at locations different from those for the radio emission.

The linear polarization angle at 22--86~GHz bands
showed no fast rotation over the time period of observations,
and remained between 22$^{\circ}$ and 50$^{\circ}$ with a median angle of $\sim36^{\circ}$.
The intrinsic polarization angles corrected for Faraday rotation
ranged from 30$^{\circ}$ to 53$^{\circ}$ with a median anlge
of $\sim42^{\circ}$,
which is aligned to the jet direction of $\sim-142^{\circ}$ observed
by the KVN at 22--129~GHz~\citep{lee+15}.
The radio polarization angle shows a behavior similar to
the optical polarization angle over MJD 56615 and 56775
as reported by \cite{hayashida+15}
who reported an optical polarization angle
around 50$^{\circ}$.
The degree of linear polarization shows no prominent change or correlation
corresponding to the $\gamma$-ray
flare,
although there were
local enhancements or a global decrease with time of $<$50\%.
These imply that
the radio and optical polarization emission regions
may be different from the region responsible for the $\gamma$-ray
flare
and are maintained by the dominant magnetic field perpendicular
to the direction of the radio jet at milliarcseconds scales.

For further investigation of the multi-frequency radio polarization of
3C~279,
the degree of linear polarization at 22--86~GHz bands was
fitted with a power law function of
$p(\%)=A\lambda^{\beta}$, suggested by
\cite{tri91} and \cite{far+14},
where $A$ is constant, $\beta$ is a polarization spectral index,
and $\lambda$ is the observing wavelength in cm.
The best fit results of the power law model to the data are shown in
Figure~\ref{fig:depolarization},
yielding $\beta=-0.01\sim-0.23$.
Since the power law model of the degree of linear polarization
explains the effect of external Faraday depolarization at
longer wavelengths~\citep[see][]{burn66},
the best fit results indicate that
the effect of external Faraday depolarization is very small
within the polarization regions corresponding to
the observing beams ($30^{''}-130^{''}$) of the KVN single-dish radio telescopes.
This result seems in contradiction to the relatively high values
of the estimated Faraday rotation RM from
$-300$ to $-1200$~rad~m$^{-2}$ between the 22~GHz and 43~GHz bands,
and from $-800$ to $-5100$~rad~m$^{-2}$ between the 43~GHz and 86~GHz bands.
However, as suggested by \cite{lee+15} for the multi-frequency polarization
at a single epoch within the period of our observations,
these results indicate that
the polarization emission of 3C~279 at 22--86~GHz bands
travels through a Faraday screen containing a uniform magnetic field
that either persisted for $\sim$400 days after the $\gamma$-ray
flare, or was not affected by the $\gamma$-ray events.

The stable, global characteristic of the magnetic field in the Faraday screen,
unaffected
by the $\gamma$-ray
flare,
suggest
that the Faraday screen may be located away from the regions
corresponding to the $\gamma$-ray
flare
so that the distribution of the dominant magnetic field
is decoupled from the $\gamma$-ray flaring mechanism.
We can constrain the location of the Faraday screen
by investigating the frequency dependence of RM through
a simple model~\citep{jor+07}.
The RM depends on the electron density $n_{\rm e}$,
the parallel magnetic field strength to the line of sight $B_{||}$,
and the path length as ${\rm RM}\propto\int n_{\rm e}B_{||}dl$.
This simple jet model leads to a RM dependence on the observing frequency
as above, assuming that, for optically thick VLBI cores,
the distance $r$ of the emission region to the central engine depends on
the observing frequency $\nu$ as $r\propto\nu^{-1}$,
the electron density $n_{\rm e}$
and the magnetic field parallel to the line of sight $B_{||}$
in the region evolve as $n_{\rm e}\propto r^{-\rm n}$
and $B_{||}\propto r^{-1}$ along the jet,
and the path length $l$ increases as a function of $r$ as $l\propto r$.
For spherical or conical geometries of the jet,
the appropriate value of the index $a=2$.
This model suggests that a Faraday screen affecting the multi-frequency
polarization emission from a compact radio source
is located in or near the jet when the frequency dependence $a$
of RM is close to 2, as discussed in \cite{jor+07}.
With the multi-frequency polarization observations, we found that
the RM frequency dependence is in the range of $a=0.7-3.4$
with a mean of 2.2, as shown in Figure~\ref{fig:rotation_measure}.
This seems similar to previous measurements
of $a=1.9-3.6$ for several AGN as presented in \cite{lee+15},
although there are some cases with $a=0.7-1.9$.
This implies that the Faraday rotation of the radio polarization
at 22--86~GHz occurs in or near the jet of 3C~279
whose geometry may be spherical or conical.

\section{Summary\label{sec:summary}}

The multi-frequency polarization monitoring observations of 3C~279
enable us to find the following:
\begin{enumerate}
\item{
The 2013 December $\gamma$-ray
flare
of 3C~279 originated in an emission region
optically thick at radio frequencies of 22-86~GHz,
or may have been generated by different populations of electrons
at locations different from those emitting the radio emission.
}
\item{
%The radio and optical polarization emission regions
%may be different from the region responsible for the $\gamma$-ray
%flare, and
The radio and optical polarization emissions originated from
different region responsible for the $\gamma$-ray emission region
are maintained by the dominant magnetic field perpendicular
to the direction of the radio jet at milliarcsecond scales.
}
\item{
The polarization emission of 3C~279 at 22-86~GHz bands
experience Faraday rotation by a Faraday screen containing a uniform magnetic field
that persisted over $\sim$400 days after the $\gamma$-ray
flare,
or was not affected by the $\gamma$-ray events.
To the extent that a simple jet model is applicable to the radio
emission from 3C~279,
we suggest that the Faraday rotation of the radio polarization
at 22-86~GHz is generated in or near the jet of 3C~279
whose geometry may be spherical or conical.
}
\end{enumerate}

%%% ACKNOWLEDGMENTS (IF ANY) %%%%%%%%%%%%%%%%%%%%%%%%%%%%%%%%%%%%%%%%

\acknowledgments

We are grateful to Sascha Trippe for important comments
and suggestions which have enormously improved the manuscript.
We thank Jan Wagner for reading and improving the manuscript.
%We are grateful to all staff members in KVN
%who helped to operate the array and to correlate the data.
The KVN is a facility operated by
the Korea Astronomy and Space Science Institute.
The KVN operations are supported
by KREONET (Korea Research Environment Open NETwork)
which is managed and operated
by KISTI (Korea Institute of Science and Technology Information).

%%% CALL LIST OF REFERENCES (natbib STYLE) %%%%%%%%%%%%%%%%%%%%%%%%%%


\begin{thebibliography}{}

%%% PUT YOUR REFERENCES HERE %%%%%%%%%%%%%%%%%%%%%%%%%%%%%%%%%%%%%%%%
\bibitem[Abdo et al.(2010)]{abd+10}
        Abdo, A. A., Ackermann, M., Ajello, M., et al.
        2010, A change in the optical polarization associated with a $\gamma$-ray flare in the blazar 3C279,
        Nature, 463, 919

\bibitem[Aleksi{\'c} et al.(2014)]{ale+14}
        Aleksi{\'c}, J., Ansoldi, S., Antonelli, L. A., et al.
        2011, MAGIC observations and multi-frequency properties of the flat spectrum radio quasar 3C~279 in 2011,
        A\&A, 567, 41

\bibitem[Aumont et al.(2010)]{aum+10}
        Aumont, J., Conversi, L., Thum, C., et al.
        2010, Measurement of the Crab Nebula Polarization at 90 GHz as a Calibrator for CMB Experiments, A\&A, 514, A70

\bibitem[Boettcher et al.(2012)]{boe+12}
        Boettcher, M., Harris, D. E., \& Krawczynski, H.
        2012, Relativistic Jets from Active Galactic Nuclei (Berlin: Wiley)

\bibitem[Burn (1966)]{burn66}
        Burn, B.~J.
        1966, On the depolarization of discrete radio sources by Faraday dispersion,
        MNRAS, 133, 67

\bibitem[Chatterjee et al.(2008)]{cha+08}
        Chatterjee, R., Jorstad, S. G., Marscher, A. P., et al.
        2008, Correlated Multi-Wave Band Variability in the Blazar 3C~279 from 1996 to 2007,
        ApJ, 689, 79

\bibitem[D'Ammando et al.(2011)]{dam+11}
        D'Ammando, F., Raiteri, C. M., Villata, M., et al.
        2011, AGILE detection of extreme $\gamma$-ray activity from the blazar PKS 1510-089 during March 2009,
        A\&A, 529, 145

\bibitem[Farnes et al.(2014)]{far+14}
        Farnes, J.~S., Gaensler, B. M., \& Carretti, E.
        2014, A Broadband Polarization Catalog of Extragalactic Radio Sources,
        ApJS, 212, 15

\bibitem[Flett \& Henderson (1979)]{fh79}
        Flett, A. M. \& Henderson, C.
        1979, Observations of the polarized emission of Taurus A, Cas A and Cygnus A at 9-mm wavelength,
        MNRAS, 189, 867


\bibitem[Hada et al.(2012)]{hada+12}
        Hada K., Kino M., Nagai H., et al. 2012, VLBI Observations of the Jet in M 87 during the Very High Energy $\gamma$-ray Flare in 2010 April, Apj, 760, 52H

\bibitem[Hayashida et al.(2015)]{hayashida+15}
        Hayashida M., Nalewajko K.,Madejski G. M., et al. 2015, Rapid Variability of Blazar 3C~279 during Flaring States in 2013-2014 with Joint {\it{Fermi}}-LAT, {\it{NuSTAR}}, {\it{SWIFT}}, and Ground-based Multi-wavelength Observations Apj, 807, 79H

\bibitem[Jorstad et al.(2007)]{jor+07}
        Jorstad, S.~G., Marscher, A.~P., Stevens, J.~A., et al.
        2007, Multiwaveband Polarimetric Observations of 15 Active Galactic Nuclei at High Frequencies: Correlated Polarization Behavior,
        AJ, 134, 799

\bibitem[Krawczynski et al.(2012)]{kra+12}
        Krawczynski, H., Boettcher, M., \& Reimer, A.
        2012, Unresolved Emission from the Core: Observations and Models,
        Relativistic Jets from Active Galactic Nuclei (Berlin: Wiley), 218

\bibitem[Krolik(1999)]{kro99}
        Krolik, J.~H.
        1999, Active galactic nuclei : from the central black hole to the galactic environment (Princeton:Princeton University Press)

\bibitem[Lee et al.(2011)]{lee+11}
        Lee, S.-S., Byun, D.-Y., Oh, C. ~S., et al. 2011, Single-Dish Performance of KVN 21-m Radio Telescopes: Simultaneous Observations at 22 and 43~GHz, PASP, 123, 1398

\bibitem[Lee et al. (2013)]{lee+13}
        Lee, S.-S., Han, M., Kang, S., et al.
        2013, Proceedings of The Innermost Regions of Relativistic Jets and Their Magnetic Fields (EPJ Web of Conferences), ed. J. L. Gomez, 61

\bibitem[Lee et al.(2014)]{lee+14}
        Lee, S.-S., Petrov, L., Byun, D.-Y., et al. 2014, Early Science with the Korean VLBI Network: Evaluation of System Performance, AJ, 147, 77

\bibitem[Lee et al.(2015)]{lee+15}
        Lee, S.-S., Kang, S., Byun, D.-Y., et al.
        2015, First Detection of 350 Micron Polarization from a Radio-loud AGN,
        ApJ, 808, 26L


\bibitem[Larionov et al.(2008)]{lar+08}
        Larionov, V. M., Jorstad, S. G., Marscher, A. P., et al.
        2008, Results of WEBT, VLBA and RXTE monitoring of 3C~279 during 2006-2007,
        A\&A, 492, 389

\bibitem[Mangum(2000)]{man00}
        Mangum, J.~G., 2000, User's Manusal for the NRAO 12 Meter Millimeter-Wave Telescope, NRAO, 129

\bibitem[Marscher et al.(2008)]{mar+08}
        Marscher, A. P., Jorstad, S. G., D'Arcangelo, F. D., et al.
        2008, The inner jet of an active galactic nucleus as revealed by a radio-to-$\gamma$-ray outburst,
        Nature, 452, 966


\bibitem[Marscher et al.(2010)]{mar+10}
        Mascher, A., Jorstad, S., Larionov, V., et al. 2010, Probing the Inner Jet of the Quasar PKS 1510-089 with Multi-waveband Monitoring During Strong Gamma-ray Activity, ApJ, 710, 126

\bibitem[Orienti et al.(2011)]{ori+11}
        Orienti, M., Venturi, T., Dallacasa, D., et al.
        2011, Multi-epoch parsec-scale observations of the blazar PKS 1510−089,
        MNRAS, 417, 359

\bibitem[Perley \& Butler(2013)]{pb13}
        Perley, R. A.; Butler, B. J.
        2013, Integrated Polarization Properties of 3C48, 3C138, 3C147, and 3C286,
        ApJS, 206, 16

\bibitem[Sasada et al.(2010)]{sas+10}
        Sasada, M., Uemura, M., Arai, A., et al.
        2010, Multiband Photopolarimetric Monitoring of an Outburst of the Blazar 3C~454.3 in 2007,
        PASJ, 62, 645

\bibitem[Sault et al.(1996)]{Sault et al. (1996)}
        Sault R. J., Hamaker J. P., and Bregman J. D. 1996, Understanding radio polarimetry. II. Instrumental calibration of an interferometer array, A\&AS, 117, 149

\bibitem[Tribble(1991)]{tri91}
        Tribble, P.~C.
        1991, Depolarization of extended radio sources by a foreground Faraday screen,
        MNRAS, 250, 726

\bibitem[Wehrle et al. (2012)]{weh+12}
        Wehrle, A. E., Marscher, A. P., Jorstad, S. G., et al.
        2012, Multiwavelength Variations of 3C~454.3 during the 2010 November to 2011 January Outburst,
        ApJ, 758, 72


%%% END LIST OF REFERENCES %%%%%%%%%%%%%%%%%%%%%%%%%%%%%%%%%%%%%%%%%%

\end{thebibliography}
\end{document}